\documentstyle[12pt
]{article}
\def\@maketitle{\newpage
 \null
 {\normalsize \tt \begin{flushright}
  \begin{tabular}[t]{l} \@date
  \end{tabular}
 \end{flushright}}
 \begin{center}
 \vskip 2em
 {\LARGE \@title \par} \vskip 1.5em {\large \lineskip .5em
 \begin{tabular}[t]{c}\@author
 \end{tabular}\par}
 \end{center}
 \par
 \vskip 1.5em}
\def\underset#1\to#2{\mathop{#2}\limits_{#1}}
\def\Ge2{\Ge^2}
\def\GL2{\GL^2}
\def\Ga{\alpha}
\def\Gb{\beta}

\def\Gd{\delta}
\def\Ge{\epsilon}
\def\Gf{\phi}
\def\Gg{\gamma}

\def\Gj{\varphi}

\def\Gm{\mu}
\def\Gn{\nu}
\def\Gp{\pi}
\def\Gq{\theta}
\def\Gr{\rho}
\def\Gs{\sigma}

\def\Gy{\psi}

\def\GD{\Delta}

\def\GG{\Gamma}
\def\GJ{\vartheta}
\def\GL{\Lambda}
\def\GP{\Pi}

\def\GS{\Sigma}
\def\pd{\partial}
\def\tr{{\rm tr}}

\def\slala{\not \hskip-2pt}

\title{Finite-Temperature and -Density QED: \\
   Schwinger-Dyson Equation \\
  in the Real-Time Formalism  \,\,
 \\
}
\author{
  {\sc Kei-ichi Kondo} \thanks{
    E-mail address: kondo@tansei.cc.u-tokyo.ac.jp}
\thanks{Supported by the Grant-in-Aid for Scientific Research
from the Ministry of Education, Science and Culture
(04740138).}
and {\sc Kazuhiro Yoshida}    \\
  \vspace{0.5cm} \\
  {\it Department of Physics,}\\
  {\it Faculty of Science,} \\
  {\it Chiba University,} \\
  {\it Inage-ku, Chiba 263, Japan}
  \vspace{1.2em} \\
}
\begin{document}
\date{
  hep-th/9304018 \\
  CHIBA-EP-67 \\
  March 1993
}
\maketitle
\begin{abstract}
Based on the real-time formalism, especially, on Thermo Field Dynamics, we
derive the Schwinger-Dyson gap equation  for the fermion propagator in QED
and Four-Fermion model at finite-temperature and -density.
We discuss some advantage of the real-time formalism in solving the
self-consistent gap equation, in comparison with the ordinary
imaginary-time formalism.
Once we specify the vertex function, we can write down the
SD equation with only continuous variables
without performing the discrete sum
over Matsubara frequencies which
cannot be performed in advance without further approximation
in the imaginary-time formalism.
By solving the SD equation obtained in this way, we find the
chiral-symmetry restoring transition at finite-temperature
and present the
associated phase diagram of strong coupling QED.
In solving the SD equation, we consider two approximations:
instantaneous-exchange and $p_0$-independent ones.
The former has a direct correspondence in the imaginary time formalism,
while the latter is a new approximation beyond the former,
since the latter is able to incorporate new thermal effects
which has been overlooked in the ordinary imaginary-time solution.
However both approximations are shown to give qualitatively the same
results on the finite-temperature phase transition.
\end{abstract}
\newpage
\section{Introduction}
Quantum chromodynamics (QCD) exhibits   very characteristic
dynamics of confinement of quarks and gluons in long distance
as well as the asymptotic freedom at short distance.
According to the standard cosmology, however, our universe ought
to have undergone the confinement-deconfinement transition of quarks and
gluons in the early stage of evolution.
This implies that the hadron phase at present should transfer into
 the quark-gluon-plasma phase,
 if the universe goes back to the past  at high-temperature and -density
\cite{kn:Hatsuda92,kn:Kapusta79,kn:Kalashnikov85,kn:KK79}
Furthermore, the quark-gluon-plasma is not merely a
byproduct of the theory in the sense that its existence will be verified
by  experiments of heavy ion collision in the following several years
\cite{kn:Hatsuda92}.
\par
At high temperature, the
effective coupling constant of QCD gets small owing to the asymptotic
freedom \cite{kn:CP75}.  Hence it is natural to consider that  the
perturbation theory can be applied to QCD at high temperature
successfully
\cite{kn:GPY81,kn:LvW87,kn:Toimela85a,kn:Kapusta89}.
However it becomes evident in the early stage of the
investigation of high-temperature QCD that infrared (IR) singularity
prevents one from calculating  higher order terms in perturbation theory
\cite{kn:Linde80,kn:KK82,kn:LB83,kn:KK85,kn:Toimela85b}.
\par
In view of this, our study aims at
performing the non-perturbative analysis on the phase transition at
finite-temperature and -density (FTD). For this we study the
self-consistent (SC) equation represented by the Schwinger-Dyson (SD)
equation and the Bethe-Salpeter (BS) equation. Such SC equations are
simultaneous non-linear integral equations and are able to incorporate
various non-perturbative effects self-consistently \cite{kn:Kondo92p}.
The concept of the SC equation is not new and indeed the SC equation is
a well-known  traditional method to solve the many body problem in
condensed matter physics and nuclear physics \cite{kn:AGD63,kn:FW71}.
Nevertheless SC equations turn out to
provide the powerful method also in   elementary particle
physics:
Recently various types of SC or gap equations
have been extensively studied to investigate the dynamical symmetry
breaking in gauge field theories and
have succeeded to reveal rich phase structure of
strong coupling gauge theories
\cite{kn:Kondo92p}.
In this paper we extend this type of research to the case of
finite-temperature and density.
\par
We desire to develop the method to study the phase transition at
finite-temperature and -density from the quantum field theoretical
point of view beyond the thermodynamic or phenomenological treatment. For
this, we need at first the formulation of FTD field theory. Traditional
Imaginary-Time Formalism (ITF) due to Matsubara \cite{kn:Matsubara} is
very efficient to develop the perturbation theory of FTD field theory.
In  ITF,  the field theory at zero-temperature can be
transferred into the FTD field theory according to the following
procedures:
in a given Feynman diagram, each integral over the temporal component of a
"fermion" loop momentum  is replaced by an infinite sum over "odd"
Matsubara frequencies according to the prescription:
\begin{equation}
 \int {dp_0 \over 2\Gp} f(p_0) \Longrightarrow
 T \sum_{n=-\infty}^{+\infty}
f(p_0 \rightarrow (2n+1)  \Gp T-i\Gm) ,
\end{equation}
where $n$ denotes the integer,  $T$ the temperature and $\Gm$ the chemical
potential. In contrast, each integral over the temporal component
of "bosonic" loop momentum is evaluated by summing over "even" Matsubara
frequencies, \begin{equation}
 \int {dk_0 \over 2\Gp} g(k_0) \Longrightarrow
 T \sum_{n=-\infty}^{+\infty}  g(k_0 \rightarrow 2n \Gp T).
\end{equation}
However  applying this method to the non-perturbative study is
sometimes either convenient nor easy.  Indeed the infinite sum is very
cumbersome particularly in solving the SC equation, because the
convergence of the infinite series is problematical and the discrete sum
cannot be performed in advance before we know the solution, in sharp
contrast with the perturbation theory where all the propagators are
bare.
\par
To write down explicitly the SC equation in the {\it closed form},
we must adopt the approximation or ansatz for the truncation of infinite
hierarchy of SC equations in general field theories.
Once such an ansatz (e.g., quenched ladder approximation in
the SD equation) is adopted, we should not make further approximation to
solve SC equations.
However this forces us to solve the infinite set of integral
equations distinguished by an integer $n$ in ITF.
In most studies based on the SC equation in IFT, therefore, infinite sum
over discrete Matsubara frequencies was evaded by searching for
frequency-independent approximate solution from the first
\cite{kn:Kalashnikov88,kn:CKV88,kn:Kocic87,kn:DM91a,kn:DM91b,kn:OS91,kn:KKK93}.
To simplify further the equation, the constant (i.e.,
momentum-independent) solution has been investigated
as in the NJL model in the leading 1/N approximation. By using
such solutions, some qualitative features of FTD phase transition have
been investigated in QED$_4$ \cite{kn:OS91,kn:KKK93},
QED$_3$ \cite{kn:Kocic87,kn:DM91a,kn:DM91b} and
QCD$_4$ \cite{kn:Kalashnikov88,kn:CKV88}.
However it is evident that in gauge theories the zero-temperature
limit of the momentum-independent approximate solution obtained in such a
way does not necessarily coincide with the zero-temperature
solution \cite{kn:Kondo92p}.
To avoid this type of discrepancy, we must obtain the momentum-dependent
solution at finite-temperature as tried in \cite{kn:ADKM92}
for QED$_3$.
\par
On the other hand,
Thermo Field Dynamics (TFD), a realization of
Real-Time Formalism (RTF) of FTD field theory
\cite{kn:DJ74}, proposed by Takahashi and
Umezawa \cite{kn:TU75,kn:UMT82} needs no discrete sum and treats only
continuous variables.
Hence TFD is well suited to extend the usual field theoretical
methods to the finite-temperature and is expected to be quite efficient
to write down and solve the SC equation.
Thanks to TFD, various methods developed so far in  solving the integral
equation can be extended to SC equations in the FTD case without much
difficulty.
Actually {\it the infinite
set of SC equations indexed by an integer in ITF  is reduced to the
corresponding single SC equation with an additional argument of continuous
variable in TFD}, as exemplified for the SD equation in this paper. Such a
viewpoint has been overlooked to the best of our knowledge.
\par
This paper is the first of a series of papers which treat the transition
of finite-temperature and density in field theories by use of
self-consistent equations based on RTF, particularly
on TFD.
In this paper, restricting to the Nambu-Jona-Lasinio
(NJL)\cite{kn:NJL}-type four-fermion model and QED,
we write down the SD equation for the fermion propagator based on TFD.
Based on the solution of the SD equation for the fermion propagator, we
show the existence of chiral-symmetry-restoring transition in QED$_4$ by
obtaining the critical line which separates the low-temperature phase
where the chiral symmetry  is spontaneously  broken from the
high-temperature phase where the chiral symmetry restores.

\par
This paper is organized as follows.
In section 2, the formalism of TFD is briefly reviewed to prepare the
necessary materials and to fix the notation.
In section 3,  we treat the simplest case of NJL-type four-fermion
model in the leading 1/N approximation, i.e., chain approximation.
Within this approximation both formalism give the same SD equation for
the fermion mass function.
In section 4, we write down the SD equation in
FTD QED based on TFD.
In section 5,  we solve the SD equation under the instantaneous-exchange
approximation.
In section 6, we consider another approximation and compare the result with
the previous one. The final section is devoted to the conclusion and
perspective.

\newpage
\section{Thermo-Field-Dynamics}
In this section we recall basic materials of TFD which are necessary to
derive the SD equation.
\subsection{scalar}
Corresponding to the lagrangian of free scalar field (at zero
temperature),
\begin{equation}
 {\cal L} = {1 \over 2} (\pd_\Gm \Gf)^2 -{1 \over 2} m_0^2 \Gf^2,
\end{equation}
the free scalar propagator at zero-temperature reads
\begin{equation}
  \Delta(k) =  {i \over k^2-m_0^2+i \Ge} .
\end{equation}
In the TFD, the free scalar propagator at finite-temperature is given by
the matrix form \cite{kn:NS84,kn:KS85a}:
\begin{equation}
  i D^{ab}(k) = U(\Gb,k) \left( \matrix{ \Delta (k) & 0 \cr
0 & \Delta^{*}(k) \cr } \right) U(\Gb,k), \ \Gb \equiv 1/T,
\end{equation}
where
\begin{equation}
 U(\Gb, k)  =  \left( \matrix{ \cosh \Gq_k & \sinh \Gq_k \cr
\sinh \Gq_k & \cosh \Gq_k \cr} \right) ,
\end{equation}
with
\footnote{The notation $A:=B$ implies that $A$ is defined by $B$.}
\begin{equation}
  \cosh \Gq_k := {1 \over \sqrt{1-e^{-\Gb|k_0|}}},
\ \sinh \Gq_k := {e^{-\Gb|k_0|/2} \over \sqrt{1-e^{-\Gb|k_0|}}}.
\end{equation}
In the zero-temperature limit $\Gb \rightarrow \infty$,
$U_{ab}(\Gb, k) \rightarrow \Gd_{ab}$ and hence
the scalar particle $\Delta(k)$ and the thermal ghost $\Delta^{*}(k)$
decouple.
\par
The explicit form of the propagator reads
\begin{equation}
  i D^{ab}(k) =
\left( \matrix{ \cosh^2 \Gq_k \Delta (k) + \sinh^2 \Gq_k \Delta^{*} (k)
& \cosh \Gq_k \sinh \Gq_k [\Delta (k) +\Delta^{*} (k)]
\cr
\cosh \Gq_k \sinh \Gq_k [\Delta (k) +\Delta^{*} (k)]
& \sinh^2 \Gq_k \Delta (k) + \cosh^2 \Gq_k \Delta^{*} (k)
\cr } \right) .
\end{equation}
Using the identities,
$
\cosh^2 \GJ_{k} - \sinh^2 \GJ_{k} =  1
$
and
$\cosh^2 \GJ_{k} + \sinh^2 \GJ_{k}
$
$
= 2 \cosh^2 \GJ_{k}  -1
$
$
=  \coth {\Gb|k_0| \over 2},
$
we obtain
\begin{eqnarray}
  i D^{11}(k)
  & &=
\cosh^2 \Gq_k \Delta (k) + \sinh^2 \Gq_k \Delta^{*} (k)
\nonumber\\
  & &=
\Delta (k) + \sinh^2 \Gq_k [\Delta (k)+\Delta^{*} (k)]
\nonumber\\
& &=  \GD(k) + 2\Gp \Gd(k^2-m_0^2) N_B(k)
\nonumber\\
& &= {i \over k^2-m_0^2+i\Ge}
+ 2\Gp \Gd(k^2-m_0^2) N_B(k)
\nonumber\\
& &= {i \over k^2-m_0^2} + \Gp \Gd(k^2-m_0^2) \coth {\Gb|k_0|
\over 2}  ,
\end{eqnarray}
where $N_B(k)$ is the Bose-Einstein distribution function:
\begin{equation}
N_B(k) := \sinh^2 \Gq_k = {1 \over \exp(\Gb|k_0|)-1}.
\end{equation}
\par
The full propagator will be assumed to be of the form
\begin{equation}
  i {\cal D}^{ab}(k) = U(\Gb,k) \left( \matrix{ {\cal D}(k) & 0 \cr
0 & {\cal D}^{*}(k) \cr } \right) U(\Gb,k),
\end{equation}
with ${\cal D}(k)$ being a complex function.
\par
As a consequence of the SD equation:
\begin{equation}
  i {\cal D}^{ab}(k) = i D^{ab}(k)
+  i D^{ac}(k) (-i \GP^{cd}) i {\cal D}^{db}(k),
\end{equation}
the self-energy matrix is written as
\footnote{This is proved if there is the spectral representation
which implies the Kubo-Martin-Schwinger (KMS) equilibrium condition
\cite{kn:SU83}.} \begin{equation}
 -i \GP^{ab}(k) = U^{-1}(\Gb, k) \left( \matrix{ -i \GP(k) & 0 \cr
0 & i \GP^{*}(k) \cr } \right) U^{-1}(\Gb, k),
\end{equation}
where each component is related as follows \cite{kn:KS85a}
\begin{eqnarray}
 \GP^{12}(k) &=& \GP^{21}(k)
 = - 2i\cosh \Gq_k \sinh \Gq_k  \Im \GP(k)
 = - i\tanh 2\Gq_k \Im \GP^{11}(k),
 \nonumber\\
 \GP^{22}(k) &=& - \GP^{11*}(k).
\end{eqnarray}
This allows us to write the function ${\cal D}(k)$ as
\begin{equation}
 {\cal D}(k) = {i \over k^2-m_0^2-\GP+i\Ge}.
\end{equation}
Hence the SD equation for the scalar propagator reads
\begin{equation}
   i{\cal D}^{-1}(k) = i \Delta^{-1}(k) - \GP(k),
\end{equation}
where the real and imaginary part is related to $\GP^{11}(p)$ as
\cite{kn:KS85a}
\begin{eqnarray}
 \Re \GP(k) &=& \Re \GP^{11}(k),
 \nonumber\\
 \Im \GP(k) &=& \Ge(k_0) \tanh[{\Gb \over 2} k_0] \Im \GP^{11}(k).
\end{eqnarray}

\subsection{fermion}
For the lagrangian of free fermion field,
\begin{equation}
 {\cal L} = \bar \Gy ( i \slala \pd - m_0) \Gy ,
\end{equation}
the free fermion propagator is obtained as
\begin{equation}
  S(p) =  {i \over \slala p -m_0+i \Ge}
= i{-\slala p +m_0 \over p^2-m_0^2+i \Ge}.
\end{equation}
Then the free fermion propagator at finite-temperature (and -density) is
given by \cite{kn:KS85a}
\begin{equation}
  i S^{ab}(k) = V(\Gb,p,\Gm) \left( \matrix{ S (p) & 0 \cr
0 & S^{*}(p) \cr } \right) V(\Gb,p,\Gm),
\end{equation}
where  $S^{*}(p)$ denotes
\begin{equation}
  S^{*}(p) =  {-i \over \slala p -m_0-i \Ge}
= -i{-\slala p +m_0 \over p^2-m_0^2-i \Ge},
\end{equation}
and
\begin{equation}
 V(\Gb, p,\Gm)  =  \left( \matrix{ \cos \Gj_{p+\Gm}
& - \Ge(p_0) e^{-\Gb \Gm/2} \sin \Gj_{p+\Gm} \cr
\Ge(p_0) e^{\Gb \Gm/2} \sin \Gj_{p+\Gm}
& \cos \Gj_{p+\Gm} \cr} \right) ,
\end{equation}
with $\Ge(p_0) := \Gq(p_0)-\Gq(-p_0)$ and
\begin{equation}
  \cos \Gj_{p+\Gm}
= {\Gq(p_0) e^{\Gb(p_0+\Gm)/4}+\Gq(-p_0) e^{-\Gb(p_0+\Gm)/4}
\over \sqrt{e^{\Gb(p_0+\Gm)/2}+e^{-\Gb(p_0+\Gm)/2}}} ,
\end{equation}
\begin{equation}
 \sin \Gj_{p+\Gm}
= {\Gq(p_0) e^{-\Gb(p_0+\Gm)/4}+\Gq(-p_0) e^{\Gb(p_0+\Gm)/4}
\over \sqrt{e^{\Gb(p_0+\Gm)/2}+e^{-\Gb(p_0+\Gm)/2}}}.
\end{equation}
In the zero-temperature limit $\Gb \rightarrow \infty$,
$V_{ab}(\Gb, p,\Gm) \rightarrow \Gd_{ab}$ and hence
the thermal part $S^{*}(p)$ is separated.
\par
Explicitly writing down the matrix element:
\begin{eqnarray}
&&  i S^{ab}(p)
\nonumber\\
&=&
\left( \matrix{ \cos^2 \Gj_{p,\Gm} S (p) - \sin^2 \Gj_{p,\Gm} S^{*} (p)
& - \Ge(p_0) e^{-\Gb \Gm/2} \cos \Gj_{p,\Gm} \sin \Gj_{p,\Gm}
 [S+S^{*}] (p)
\cr
\Ge(p_0) e^{\Gb \Gm/2} \cos \Gj_{p,\Gm} \sin \Gj_{p,\Gm}
 [S+S^{*}] (p)
& - \sin^2 \Gj_{p,\Gm} S (p) + \cos^2 \Gj_{p,\Gm} S^{*} (p)
\cr } \right) .
\end{eqnarray}
Note that, since
$
\cos^2 \Gj_{p+\Gm} + \sin^2 \Gj_{p+\Gm} =  1,
$
and
$
\cos^2 \Gj_{p+\Gm} - \sin^2 \Gj_{p+\Gm}
= 2 \cos^2 \Gj_{p+\Gm}  -1
=  \Ge(p_0) \tanh {\Gb(p_0+\Gm) \over 2} ,
$
the 1-1 component of the fermion propagator is rewritten as
\begin{eqnarray}
  i S^{11}(p)
&&= \cos^2 \Gj_{p+\Gm} S (p) - \sin^2 \Gj_{p+\Gm} S^{*} (p)
\nonumber\\
&&=  S (p) - \sin^2 \Gj_{p+\Gm} [S (p) +S^{*} (p)]
\nonumber\\
&&= S(p) -2 \Gp \Ge(p_0) \Gd(p^2-m_0^2) (\slala p+m_0) N_F(p)
\nonumber\\
& &= {i \over \slala p-m_0+i\Ge}
- 2\Gp \Ge(p_0) \Gd(p^2-m_0^2) (\slala p+m_0) N_F(p)
\nonumber\\
&&=  {i \over \slala p-m_0} + \Gp \Ge(p_0) \Gd(p^2-m_0^2) (\slala p+m_0)
 \tanh {\Gb(p_0+\Gm) \over 2},
\end{eqnarray}
where $N_F(p)$ is the Fermi-Dirac distribution function:
\begin{equation}
 N_F(p) := \sin^2 \Gj_{p+\Gm}
 = {1 \over e^{\Gb(p_0+\Gm)}+1} \Gq(p_0)
+ {1 \over e^{-\Gb(p_0+\Gm)}+1} \Gq(-p_0).
\end{equation}
\par
The full propagator will have the form \cite{kn:SU83}
\begin{equation}
  i {\cal S}^{ab}(p) = V(\Gb,p) \left( \matrix{ {\cal S}(p) & 0 \cr
0 & {\cal S}^{*}(p) \cr } \right) V(\Gb,p),
\end{equation}
with ${\cal S}(p)$ being a complex function.
\par
Similarly in the case of the scalar field,
the SD equation:
\begin{equation}
  i {\cal S}^{ab}(p) = i S^{ab}(p)
+  i S^{ac}(p) (-i \GS^{cd}) i {\cal S}^{db}(p),
\end{equation}
is compatible with the self-energy matrix written as
\begin{equation}
 -i \GS^{ab}(p) = V^{-1}(\Gb, p) \left( \matrix{ -i \GS(p) & 0 \cr
0 & i \GS^{*}(p) \cr } \right) V^{-1}(\Gb, p),
\end{equation}
where
\begin{eqnarray}
 \GS^{12}(p) &=& - e^{-\Gb \Gm} \GS^{21}(p)
 = i\Ge(p_0) e^{-\Gb \Gm/2} \tan 2\Gj_{p+\Gm} \Im \GS^{11}(p),
 \nonumber\\
 \GS^{22}(p) &=& - \GS^{11*}(p).
\end{eqnarray}

This allows us to write the function ${\cal S}(p)$ as
\begin{equation}
 {\cal S}(p) = {i \over \Gg_\Gm p_\Gm-m_0-\GS(p)+i\Ge}.
\end{equation}
Hence the SD equation for the fermion propagator takes the form:
\begin{equation}
 i{\cal S}^{-1}(p) = iS^{-1}(p)-\GS(p),
\end{equation}
where
\begin{eqnarray}
 \Re \GS(p) &=& \Re \GS^{11}(p),
 \nonumber\\
 \Im \GS(p) &=& \Ge(p_0) \coth[{\Gb \over 2}(p_0+\Gm)] \Im \GS^{11}(p).
\end{eqnarray}

\subsection{photon}
Given the lagrangian for the photon field
\begin{eqnarray}
 {\cal L} = - {1 \over 4} F^{\Gm \Gn}  F_{\Gm \Gn}
 - {1 \over 2\Ga} (\pd^\Gm A_\Gm)^2,
\end{eqnarray}
with $\Ga$ being the gauge-fixing parameter, the photon propagator is
given by
\begin{equation}
 i D_{\Gm \Gn}^{ab}(k) = \left[-g_{\Gm \Gn} -(1-\Ga)
 k_{\Gm} k_{\Gn} {\pd \over \pd k^2} \right] i D^{ab}(k)|_{m=0},
\end{equation}
where $D^{ab}(k)|_{m=0}$ is the massless scalar propagator.

\par
For the full photon propagator ${\cal D}_{\Gm \Gn}^{ab}(k)$ obeying the
SD equation
\begin{equation}
  i {\cal D}_{\Gm \Gn}^{ab}(k) = i D_{\Gm \Gn}^{ab}(k)
+  i D_{\Gm \Gr}^{ac}(k) (-i \GP_{\Gr \Gs}^{cd}) i {\cal D}_{\Gs
\Gn}^{db}(k),
\end{equation}
the vacuum polarization function is introduced
\begin{equation}
 i \GP_{\Gm \Gn}^{ab}(k)
 = U^{-1}(\Gb, k) \left( \matrix{-i \GP_{\Gm \Gn}(k) & 0 \cr
0 & -i \GP_{\Gm \Gn}^{*}(k) \cr } \right) U^{-1}(\Gb, k),
\end{equation}
where
\begin{eqnarray}
 \GP_{\Gm \Gn}^{12}(k) &=& \GP_{\Gm \Gn}^{21}(k)
 = - 2i\cosh \Gq_k \sinh \Gq_k  \Im \GP_{\Gm \Gn}(k)
 = - i\tanh 2\Gq_k \Im \GP_{\Gm \Gn}^{11}(k),
 \nonumber\\
 \GP_{\Gm \Gn}^{22}(k) &=& - \GP_{\Gm \Gn}^{11*}(k).
\end{eqnarray}
It is expected to satisfy the relation
\begin{equation}
   i{\cal D}_{\Gm \Gn}^{-1}(k)
   = i D_{\Gm \Gn}^{-1}(k) - \GP_{\Gm \Gn}(k),
\end{equation}
where
\begin{eqnarray}
 \Re \GP_{\Gm \Gn}(k) &=& \Re \GP_{\Gm \Gn}^{11}(k),
 \nonumber\\
 \Im \GP_{\Gm \Gn}(k) &=& \Ge(k_0) \tanh[{\Gb \over 2} k_0]
 \Im \GP_{\Gm \Gn}^{11}(k).
\end{eqnarray}

\subsection{remark on the full propagator}
In this paper we discuss only the real part of the self-energy function
as explained in section 4.

In the general case of $\Im \GS(p)\not=0$, the full
boson propagator should take the form
\begin{equation}
i{\cal D}^{11}(k) = {i \over k^2 - \GP(k)+i\Ge}
+\left[ {i \over k^2 - \GP(k)+i\Ge}
-{i \over k^2 - \GP^{*}(k)-i\Ge} \right] N_B(k),
\end{equation}
and the full fermion propagator
\begin{equation}
i{\cal S}^{11}(p) = {i \over \slala p - \GS(p)+i\Ge}
-\left[ {i \over \slala p - \GS(p)+i\Ge}
-{i \over \slala p - \GS^{*}(p)-i\Ge} \right] N_F(p).
\end{equation}

\newpage
\section{SD equation for NJL model}
 We consider the model with four-fermion interaction of
 Nambu-Jona-Lasinio (NJL) type \cite{kn:NJL}
 whose lagrangian is given by
\begin{equation}
  {\cal L}  =  \bar \Gy^a i \slala \pd \Gy^a - m_0  \bar \Gy^a  \Gy^a
  + {1 \over 2}{G \over N}
  [(\bar \Gy^a \Gy^a)^2+(\bar \Gy^a i \Gg_5\Gy^a)^2],
  \ (a=1,...,N)
\end{equation}
which is equivalent to
\begin{equation}
  {\cal L}  =  \bar \Gy^a i \slala \pd \Gy^a
  - \bar \Gy^a(\Gs+i\Gg_5 \Gp) \Gy^a
  - {N \over G}[{1 \over 2}(\Gs^2+\Gp^2)-m_0 \Gs],
\end{equation}
where we have introduced the scalar and the pseudoscalar auxiliary fields
$\Gs$ and $\Gp$ respectively.

\par
Due to the Yukawa interaction of the fermion with the auxiliary
scalar field, two tadpole diagrams contribute to the fermion self-energy
in  the NJL model in the leading order of 1/N expansion (Fig.1), which is
the chain approximation:
\begin{equation}
  \GS^{11} = - \int {d^D p \over (2\Gp)^D}
 \{D^{11}(0) \tr[i{\cal S}^{11} (p)]
 +D^{12}(0) \tr[i{\cal S}^{22} (p)] \},
\end{equation}
where $D^{ab}(k)$ is the auxiliary field propagator.
Since  $D^{11}(0)=G/N$, $D^{12}(0)=0$,
it turns out that the second diagram has vanishing contribution.

When $m_0=0$, we notice that
\begin{equation}
  \tr[i S^{11}(p)] = i \tr(1) \left[
{M \over p^2-M^2}
+ i \Ge(p_0)\Gp \Gd(p^2-M^2)  M \tanh {\Gb(p_0+\Gm) \over 2} \right].
\end{equation}
Hence we obtain
\begin{eqnarray}
 D^{11}(0) \Re \tr[i{\cal S}^{11} (p)]
&=&  - \Gp G \tr(1) \Ge(p_0) \Gd(p_0^2-E_P^2)
 M \tanh {\Gb(p_0+\Gm) \over 2}
 \nonumber\\
&=&  - \Gp G \tr(1)
{\Gd(p_0-E_P)-\Gd(p_0+E_P) \over 2E_P}
 M \tanh {\Gb(p_0+\Gm) \over 2},
\end{eqnarray}
where
\begin{equation}
\ E_P := \sqrt{P^2 + M^2}, \ P = |\vec P|.
\end{equation}
Then the integration with respect to the time-component is
straightforward:  \begin{equation}
 \int {dp_0 \over 2\Gp}  D^{11}(0) \Re \tr[i{\cal S}^{11} (p)]
= - G {\tr(1) \over 4} {M \over E_P}
\left[\tanh {\Gb(E_P+\Gm) \over 2}+\tanh {\Gb(E_P-\Gm) \over 2}\right]  .
\end{equation}

Identifying the fermion mass $M$ with the real part of $\GS$
\footnote{See the next section for the reason.}:
\begin{equation}
M = \tr(\Re \GS)/\tr(1) = \tr(\Re \GS^{11})/\tr(1),
\end{equation}
therefore, the gap equation in the NJL$_D$ ($D \ge 2$) model
at FTD is obtained:
\begin{equation}
M = {\tr(1) \over 4} G \int {d^{D-1} P \over (2\Gp)^{D-1}}
 {M \over E_P} \left[ \tanh {\Gb(E_P+\Gm) \over 2}+
\tanh {\Gb(E_P-\Gm) \over 2} \right].
\end{equation}
The solution obtained from this gap equation is momentum-independent.
\par
This gap equation indeed coincides with that obtained by performing  the
sum over Matsubara frequencies in ITF.
\begin{eqnarray}
 \GS = - T \sum_{n=-\infty}^{n=+\infty}
 \int {d^{D-1} P \over (2\Gp)^{D-1}} \tr[S(p)]  D(0),
\end{eqnarray}
where
\begin{eqnarray}
 S(p) &=& {1 \over \slala p -\GS} = {\slala p + \GS \over p^2-\GS^2},
 \ p_0 = (2n+1)i\Gp T +\Gm,
\nonumber\\
D(0) &=&  {G \over N}.
\end{eqnarray}
The self-consistent solution $\GS$ is $p$-independent and hence
we can perform the discrete sum:
\begin{eqnarray}
 \GS &=&   \tr(1) G T \sum_{n=-\infty}^{n=+\infty}
 \int {d^{D-1} P \over (2\Gp)^{D-1}}
 { \GS \over p_0^2-E_P^2}
\nonumber\\
&=&    \tr(1) G
 \int {d^{D-1} P \over (2\Gp)^{D-1}}
 {\GS \over 4E_P}
\left[ \tanh {\Gb(E_P+\Gm) \over 2}+
\tanh {\Gb(E_P-\Gm) \over 2} \right],
\end{eqnarray}
with
$
\ E_P := \sqrt{P^2 + \GS^2}.
$

\newpage
\section{QED at finite-temperature and -density}

It should be mentioned on treatment of imaginary part of
self-energy function in this paper.
In TFD, two types of field indexed by 1 and 2 appear in the theory
where the type-1 field is the usual field and the type-2 newly
introduced tilde field corresponding to the ghost field in the heat bath
\cite{kn:NS84}.
\par
The self-energy function $\GS(p)$ differs from $\GS^{11}(p)$, although the
real parts  of $\GS(p)$ and $\GS^{11}(p)$ coincide \cite{kn:KS85a}:
\begin{equation}
  \Re \GS(p) = \Re \GS^{11}(p) ,
  \  \Im \GS(p) = \Ge(p_0) \coth {\Gb(p_0+\Gm) \over 2} \Im \GS^{11}(p).
\end{equation}
In this paper we consider only the real part of the self-energy
$\Re \GS(p)$. In other
words, we search for a self-consistent solution $\GS(p)$ satisfying
$\Im \GS(p) \equiv 0$:
\begin{equation}
 \Re \GS(p) \equiv \Re \GS^{11}(p), \ \Im \GS(p) \equiv 0.
\end{equation}
This is one possible solution in the scheme of SC equations.
This standpoint is different from the  finite-temperature perturbation
theory \cite{kn:KS85a,kn:KS85b}. An imaginary part of the
self-energy at finite temperature describes the approach to equilibrium
and can be related to the dissipative transport coefficients of viscosity
and heat conductivity \cite{kn:KS85b}.
In this paper we consider only the equilibrium
case and neglect the imaginary part which will be discussed in the
subsequent paper. \footnote{The limit $\Im \GS(p)\rightarrow 0$ should be
taken from the negative side $\Im \GS(p) < 0$.}

\subsection{derivation of the SD equation from TFD}
We consider the fermion self-energy function $\GS^{11}(p)$ in QED in
$D$-dimensions (QED$_D$).
In the bare vertex approximation, the first non-trivial contribution to
$\GS^{11}(p)$ is written by using only
 ${\cal S}^{11}$ and ${\cal D}_{\Gm \Gn}^{11}$ as
diagrammatically shown in Fig.2:
\begin{equation}
  \GS^{11}(p) = e_0^2 \int {d^D q \over (2\Gp)^D}
\int {d^D k \over (2\Gp)^D} (2\Gp)^D \Gd(p-q-k)
\Gg_\Gm i{\cal S}^{11}(q) \Gg_\Gn  i{\cal D}_{\Gm \Gn}^{11}(k),
\end{equation}
since the 1-1 component ${\cal S}^{11}$ of the full fermion propagator
corresponds to
two physical type-1 external legs.
\par
In this paper, moreover, we search for the self-consistent (real)
solution of the full fermion propagator in the following form:
\begin{equation}
i{\cal S}^{11}(p) = Z(p) \left[{i \over \slala p - M(p)}
+ \Gp \Ge(p_0) \Gd(p^2-M^2(p))
\tanh {\Gb(p_0+\Gm) \over 2} (\slala p+M(p)) \right] ,
\end{equation}
and we choose the  photon propagator in the Feynman gauge
\footnote{
At zero temperature, the bare vertex approximation and the free photon
propagator in the Landau gauge leads to no wave function renormalization
of the fermion propagator, $Z(p) \equiv 1$,  \cite{kn:Kondo92p} which is
consistent with the bare vertex approximation in light of the Ward
identity, $Z_1=Z_2$.
This property is not preserved at finite temperature in the exact sense.
However the
deviation of $Z(p)$ from $1$ is at most logarithmic in the momentum and
does not substantially change the result in this paper.
This will be discussed in detail in the subsequent paper.}
at finite-temperature:
\begin{equation}
i{\cal D}_{\Gm \Gn}^{11}(k)
= - g_{\Gm \Gn}\left[{i \over k^2-\GP(k)}
+\Gp \Gd(k^2-\GP(k)) \coth {\Gb|k_0| \over 2} \right]
= - g_{\Gm \Gn} i{\cal D}^{11}(k),
\end{equation}
where $\GP(k)$ is the vacuum polarization function of the photon.
Then the real part of $\GS^{11}(p)$ is given by
\begin{eqnarray}
  \Re \GS^{11}(p)
= && -e_0^2  \Gp   \int {d^D q \over (2\Gp)^D}
\int {d^D k \over (2\Gp)^D}
(2\Gp)^D \Gd(p-q-k)  Z(q)
\Gg_\Gm (\slala q+M(q)) \Gg_\Gm
\nonumber\\
&&    \Big[ {\Ge(q_0) \Gd(q^2-M^2(q))
\tanh {\Gb(q_0+\Gm) \over 2} \over k^2-\GP(k)}
+ {\Gd(k^2-\GP(k)) \coth {\Gb|k_0| \over 2} \over q^2-M^2(q)}
 \Big].
\end{eqnarray}
Here we define the wave function
renormalization function $Z(p)$ and the fermion mass function $M(p)$
through
\begin{eqnarray}
 {\cal S}(p) = {i \over \slala p - \GS(p)}
 = {i Z(p) \over \slala p - M(p)}.
\end{eqnarray}
Hence we obtain
\begin{eqnarray}
 M(p) &=& {\tr[\Re \GS(p)] \over \tr(1)}
 = {\tr[\Re \GS^{11}(p)] \over \tr(1)},
\nonumber\\
Z^{-1}(p) &=& 1-{\tr[\slala p \Re \GS(p)] \over p^2\tr(1)}
 =1-{\tr[\slala p \Re \GS^{11}(p)] \over p^2\tr(1)}.
\end{eqnarray}
Thus we can write down the coupled SD equation
in FTD QED$_D$ as
\begin{eqnarray}
Z^{-1}(p) =  1 - (2-D) e_0^2  \Gp    \int {d^D q \over (2\Gp)^D}
{p \cdot q \over p^2} Z(q)  &\Big[&
    {\Ge(q_0) \Gd(q^2-M^2(q))
\tanh [{\Gb \over 2}(q_0+\Gm)] \over
(p-q)^2-\GP(p-q)}
\nonumber\\
&+&  {\Gd((p-q)^2-\GP(p-q)) \coth [{\Gb \over 2}|p_0-q_0|]
\over q^2-M^2(q)}
 \Big],
\end{eqnarray}
\begin{eqnarray}
{M(p) \over Z(p)} =   - D e_0^2  \Gp  \int {d^D q \over (2\Gp)^D}
Z(q)M(q) &\Big[&
    {\Ge(q_0) \Gd(q^2-M^2(q))
\tanh [{\Gb \over 2}(q_0+\Gm)] \over
(p-q)^2-\GP(p-q)}
\nonumber\\
&+&  {\Gd((p-q)^2-\GP(p-q)) \coth [{\Gb \over 2}|p_0-q_0|]
\over q^2-M^2(q)}
 \Big].
\end{eqnarray}
It is easy to see that if $M(p)$ is a solution of the SD equation, then
$-M(p)$ is also a solution.
\par
In ITF we must solve infinite number of
integral equations (indexed by an integer) for the function $M_n(P)$ with
one argument, see Section 5.1. In the TFD, on the other hand, we have only
to solve the single integral equation for the function $M(p_0,P)$ with two
arguments. \par

\subsection{finite-temperature and -density QED$_4$}
In what follows, we put $Z(p) \equiv 1$ as explained in the footnote.
For convenience, we decompose the self-energy function
$\GS(p_0,P;T,\Gm)$, right-hand-side of the above equation, into two
parts, $\GS_f(p_0,P;T,\Gm)$ and $\GS_{ph}(p_0,P;T)$.
Thus the SD equation for the fermion mass function
$M(p_0,P)$ is written as
\begin{equation}
 M(p_0,P) = \GS_f(p_0,P;T,\Gm) + \GS_{ph}(p_0,P;T).
\end{equation}
In what follows, we replace for simplicity the vacuum polarization
function of the photon with its infrared value, i.e., mass of the photon
generated by finite-temperature and -density effects
\cite{kn:Weldon82}.
\begin{equation}
 \GP(k) \Longrightarrow \GP(0) \equiv m^2 = m^2(T,\Gm).
\end{equation}
Improvement of this approximation will be tackled in the subsequent
paper.
\par
The first self-energy part reads
\begin{eqnarray}
\GS_f(p_0,P;T,\Gm)
=   - e_0^2  \Gp  D  \int {d^D q \over (2\Gp)^D}
M(q_0,Q)
   {\Ge(q_0) \Gd(q^2-M^2(q))
\tanh {\Gb(q_0+\Gm) \over 2} \over (p_0-q_0)^2-(\vec P-\vec Q)^2-m^2}.
\end{eqnarray}
Decomposing the integration measure into the angular part
and the radial one, we obtain for
\footnote{$D=3$ case (QED$_3$) will be discussed in a separate paper.}
$D >3$
\begin{eqnarray}
\GS_f(p_0,P;T,\Gm)
=  - && e_0^2  \Gp  D C_D \int_{-\infty}^{+\infty} {d q_0 \over 2\Gp}
\int_0^\GL Q^{D-2}dQ \int_0^\Gp d\GJ \sin ^{D-3} \GJ
\nonumber\\
&&  M(q_0,Q)
   {\Ge(q_0) \Gd(q^2-M^2(q))
\tanh {\Gb(q_0+\Gm) \over 2} \over
(p_0-q_0)^2-(P^2+Q^2+m^2)+2PQ \cos \GJ},
\end{eqnarray}
where $P=|\vec P|$, $Q=|\vec Q|$, $\cos \GJ=\vec P \cdot \vec Q/PQ$ and
\begin{eqnarray}
C_D
= {2 \over (2\Gp)^{D-1}} {\Gp^{{D-2 \over 2}} \over \GG({D-2 \over 2})}
= {1 \over 2^{D-2} \Gp^{D/2} \GG({D-2 \over 2})}, \ (D > 3).
\end{eqnarray}
Here we have introduced the ultraviolet (UV) momentum cutoff $\GL$.
\par
For $D=4$, after performing the angular integration, we obtain
\begin{eqnarray}
\GS_f(p_0,P;T, \Gm)
=  &&   {\Ga \over \Gp} \int_{-\infty}^{+\infty}  d q_0
\int_0^\GL dQ  \Ge(q_0) \Gd(q^2-M^2(q))
\nonumber\\
&& M(q_0,Q)
\tanh [{\Gb \over 2}(q_0+\Gm)]
{Q \over P} \ln {(p_0-q_0)^2-(P+Q)^2-m^2 \over (p_0-q_0)^2-(P-Q)^2-m^2},
\end{eqnarray}
where we have introduced a new coupling constant ($C_4=1/(4\Gp^2)$)
\begin{eqnarray}
 \Ga := {e_0^2 \over 4\Gp}.
\end{eqnarray}
Next we carry out the integration with respect to $q_0$ in the formal
way.
Assuming that the equation
\begin{equation}
  p_0^2-P^2-M^2(p_0,P) = 0
\end{equation}
has two solutions:
\footnote{We assume that such a pair of solutions exists.
Note that invariance of the mass function
 $M(p_0, P) = M(-p_0,P)$ under the reflection
$p_0 \rightarrow -p_0$ does not hold in the presence of the chemical
potential $\Gm\not=0$.
If the solution has no $p_0$-dependence,
$M(p_0,P)= \tilde M(P)$, then $p_0=\pm E_P$,
$E_P=\sqrt{P^2+\tilde M^2(P)}$.}
$p_0=\pm E_P^{\pm}$ $(E_P^{\pm}>0)$,
\begin{eqnarray}
&&  \Gd(p^2-M^2(p_0,P))
\nonumber\\
&&=  \Gd(p_0^2-P^2-M^2(p_0,P))
\nonumber\\
&&=  {\Gd(p_0-E_P^+)  \over
|2E_P^+ - {d \over dp_0}M^2(p_0,P)|_{p_0=+E_P^+}}
+  {\Gd(p_0+E_P^-) \over
|-2E_P^- - {d \over dp_0}M^2(p_0,P)|_{p_0=-E_P^-}}.
\end{eqnarray}

Using this formula, the first self-energy part $\GS_f(p_0,P;T,\Gm)$
reads
\begin{eqnarray}
\GS_f(p_0,P;T,\Gm)
&=&  {\Ga \over \Gp}  \int_0^\GL   d Q
{Q  \over P}
\nonumber\\
&& \Big[  {M(E_Q^+,Q) \tanh {\Gb(E_Q+\Gm) \over 2}
\over |2E_Q^+ - {d \over dq_0}M^2(q_0,Q)|_{q_0=+E_Q^+}}
 \ln {(p_0-E_Q)^2-(E_m^{+})^2 \over (p_0-E_Q)^2-(E_m^{-})^2}
\nonumber\\
&& + {M(-E_Q^-,Q) \tanh {\Gb(E_Q-\Gm) \over 2}
\over |-2E_Q^- - {d \over dq_0}M^2(q_0,Q)|_{q_0=-E_Q^-}}
 \ln {(p_0+E_Q)^2-(E_m^{+})^2 \over (p_0+E_Q)^2-(E_m^{-})^2}
\Big],
\end{eqnarray}
where
\begin{eqnarray}
 E_m^{\pm} := \sqrt{| P\pm Q|^2+m^2} .
\end{eqnarray}
\par
On the other hand, the second self-energy part $\GS_{ph}(p_0,P;T)$ reads
$(D>3)$
\begin{eqnarray}
\GS_{ph}(p_0,P;T)
&=&  -   \Gp  D e_0^2    \int {d^D q \over (2\Gp)^D}
M(q_0,Q)  {\Gd((p-q)^2-m^2) \coth {\Gb|(p-q)_0| \over 2}
\over q_0^2-Q^2-M^2(q_0,Q)}
\nonumber\\
&=&  -  \Gp D C_D e_0^2    \int_{-\infty}^{+\infty} {d q_0 \over 2\Gp}
\int_{0}^{\GL} Q^{D-2} dQ \int_0^\Gp d\GJ \sin^{D-3}\GJ
  {M(q_0,Q)\coth {\Gb|(p-q)_0| \over 2}
\over q_0^2-Q^2-M^2(q_0,Q)}
\nonumber\\
&&\Gd((p_0-q_0)^2-(P^2+Q^2+m^2)+2PQ \cos \GJ).
\end{eqnarray}
Note that $\GS_{ph}(p_0,P;T)$ has no $\Gm$-dependence.
\par
For $D=4$, after performing the angular integration, we obtain
\begin{eqnarray}
\GS_{ph}(p_0,P;T)
 =   - {\Ga \over \Gp}
\int_0^\GL  dQ \int_{{\cal D}} d q_0  {Q \over P}
  {M(q_0,Q)\coth {\Gb|(p-q)_0| \over 2}
\over q_0^2-Q^2-M^2(q_0,Q)}.
\end{eqnarray}
Note that the range of
integration $-1 < \cos \GJ \le 1$ is equivalent to
${\cal D}= \{\sqrt{|P-Q|^2+m^2}$ $\le |p_0-q_0| $$< \sqrt{(P+Q)^2+m^2}\}$
from the fact that the delta function has the support at a  point
$\cos \GJ={P^2+Q^2+m^2-(p_0-q_0)^2 \over 2PQ}$.
Then we obtain
\begin{eqnarray}
\GS_{ph}(p_0,P;T)
=    {\Ga \over \Gp} \int_0^\GL  dQ {Q \over P} &\Biggr[&
 \int_{p_0-E_m^{-}}^{p_0-E_m^{+}}
d q_0   M(q_0,Q)
{\coth {\Gb(p-q)_0 \over 2} \over q_0^2-Q^2-M^2(q_0,Q)}
\nonumber\\
&-&  \int_{p_0+E_m^{-}}^{p_0+E_m^{+}}
d q_0   M(q_0,Q)
{\coth {\Gb(q-p)_0 \over 2} \over q_0^2-Q^2-M^2(q_0,Q)}
\Biggr] .
\end{eqnarray}
Integration with respect to $q_0$ can not be performed for
$\GS_{ph}(p_0,P;T)$, unless we
know the dependence of the solution $M(q_0,Q)$ on $q_0$.
\par
Thus the SD equation of FTD QED$_4$ for the fermion mass function is
written as
\begin{eqnarray}
M(p_0,P)
&=&  {\Ga \over \Gp}  \int_0^\GL   d Q
{Q  \over P}
 \Biggr[  {M(E_Q^+,Q) \tanh {\Gb(E_Q+\Gm) \over 2}
\over |2E_Q^+ - {d \over dq_0}M^2(q_0,Q)|_{q_0=+E_Q^+}}
 \ln {(p_0-E_Q)^2-(E_m^{+})^2 \over (p_0-E_Q)^2-(E_m^{-})^2}
\nonumber\\
&& + {M(-E_Q^-,Q) \tanh {\Gb(E_Q-\Gm) \over 2}
\over |-2E_Q^- - {d \over dq_0}M^2(q_0,Q)|_{q_0=-E_Q^-}}
 \ln {(p_0+E_Q)^2-(E_m^{+})^2 \over (p_0+E_Q)^2-(E_m^{-})^2}
\Biggr]
\nonumber\\
&&+    {\Ga \over \Gp} \int_0^\GL  dQ {Q \over P} \Biggr[
 \int_{p_0-E_m^{-}}^{p_0-E_m^{+}}
d q_0   M(q_0,Q)
{\coth {\Gb(p-q)_0 \over 2} \over q_0^2-Q^2-M^2(q_0,Q)}
\nonumber\\
&&-  \int_{p_0+E_m^{-}}^{p_0+E_m^{+}}
d q_0   M(q_0,Q)
{\coth {\Gb(q-p)_0 \over 2} \over q_0^2-Q^2-M^2(q_0,Q)}
\Biggr] .
\end{eqnarray}

\par

\subsection{limiting cases of QED$_4$}
The SD equation in the various limiting cases is obtained in the
following replacements.
In the zero-temperature limit $\Gb \rightarrow
\infty$,  \begin{eqnarray}
&& \tanh {\Gb(E_Q-\Gm) \over 2}
\rightarrow \Gq(E_Q-\Gm)-\Gq(\Gm-E_Q),
\nonumber\\
&& \tanh {\Gb(E_Q+\Gm) \over 2}
\rightarrow 1 \equiv \Gq(E_Q-\Gm)+\Gq(\Gm-E_Q),
\nonumber\\
&& \coth {\Gb|p_0-q_0| \over 2} \rightarrow 1 .
\end{eqnarray}
In the zero-temperature $\Gb \rightarrow \infty$
and the zero-density limit $\Gm \rightarrow 0$,
\begin{eqnarray}
&& \tanh {\Gb(E_Q-\Gm) \over 2} ,
 \tanh {\Gb(E_Q+\Gm) \over 2} \rightarrow 1 ,
\nonumber\\
&& \coth {\Gb|p_0-q_0| \over 2} \rightarrow 1 .
\end{eqnarray}


\newpage
\section{Instantaneous-exchange approximation}
\subsection{instantaneous-exchange approximation in ITF}
In the imaginary-time formalism, the full fermion propagator as the
solution of the SD equation has the form (when the wave function
renormalization can be neglected):
\begin{eqnarray}
{\cal S}^{-1}(p)=\slala p+ \GS_m(P), \ p_0=(2m+1)\Gp T-i\Gm,
\end{eqnarray}
which becomes frequency-dependent.
Then we can not
perform the discrete sum in advance,
before we find the solution of the self-consistent SD equation:
\begin{eqnarray}
 \GS_m(P) =  e_0^2 T
\sum_{n=-\infty}^{+\infty} \int {d^{D-1}Q \over (2\Gp)^{D-1}}
 D_{\Gm \Gm}(k_0=p_0-q_0,\vec K=\vec P-\vec Q)
 {\GS_n(Q) \over q_0^2 + Q^2 + \GS_n^2(Q)},
\end{eqnarray}
with $q_0=(2n+1)\Gp T-i\Gm$.
This fact
renders the analytical treatment as well as the numerical one extremely
cumbersome.
\par
To avoid this problem,
the  {\sl instantaneous-exchange (IE) approximation}\cite{kn:DM91a}
is adopted:
\begin{eqnarray}
 D_{\Gm \Gm}(k_0,\vec K) \simeq  D_{\Gm \Gm}(k_0=0,\vec K).
\end{eqnarray}
In this approximation the solution $\GS_m(P)$ becomes
frequency-independent  $\GS(P)$ and the
summation over discrete frequencies can be performed explicitly
as
\begin{eqnarray}
 \GS(P) =  e_0^2
 \int {d^{D-1}Q \over (2\Gp)^{D-1}}
 D_{\Gm \Gm}(0,\vec P-\vec Q)
 {\GS(Q) \over 4E_Q}
 \left[  \tanh {\Gb(E_Q+\Gm) \over 2}
+  \tanh {\Gb(E_Q-\Gm) \over 2} \right],
\end{eqnarray}
where
\begin{eqnarray}
 E_Q = \sqrt{Q^2+\GS^2(Q)}.
\end{eqnarray}
For our choice
\begin{eqnarray}
 D_{\Gm \Gn}(0,\vec K=\vec P-\vec Q)
= {g_{\Gm \Gn} \over (\vec P-\vec Q)^2+m^2},
\end{eqnarray}
the angular integration is performed
for $D > 3$ as
\begin{eqnarray}
 \GS(P) =  D C_D e_0^2
 \int_0^\GL Q^{D-2}dQ  {\GS(Q) \over 4E_Q}
 \left[  \tanh {\Gb(E_Q+\Gm) \over 2}
+  \tanh {\Gb(E_Q-\Gm) \over 2} \right]
\nonumber\\
 \times \int_0^\Gp d\GJ
 {\sin ^{D-3} \GJ \over P^2+Q^2+m^2-2PQ \cos \GJ}.
\end{eqnarray}
Then the SD equation for QED$_4$ in IE approximation is obtained
\begin{eqnarray}
 \GS(P) =  {\Ga \over \Gp}
 \int_0^\GL dQ
 {Q \GS(Q) \over PE_Q}
 \left[  \tanh {\Gb(E_Q+\Gm) \over 2}
+  \tanh {\Gb(E_Q-\Gm) \over 2} \right]
\ln {(P+Q)^2+m^2 \over (P-Q)^2+m^2}.
\end{eqnarray}
\par
\subsection{instantaneous-exchange approximation in TFD}
 We consider the similar approximation of the SD equation
in TFD, which we call also the IE approximation.
If we put $k_0=0$ in the photon propagator,
$i{\cal D}_{\Gm \Gn}^{11}(k_0, \vec K),$ the temperature-dependence
piece   vanishes from the nature of the delta-function
$\Gd(-\vec K^2-m^2)=0$
as long as $m^2>0$, i.e.,
\begin{equation}
i{\cal D}_{\Gm \Gn}^{11}(k_0=0, \vec K)
= - g_{\Gm \Gn}{i \over -\vec K^2-m^2}.
\end{equation}
Therefore, in our IE approximation
\begin{equation}
\GS_{ph}(p_0,P;T) \equiv 0.
\end{equation}
Furthermore,  $\GS_{f}(p_0,P;T,\Gm)$ becomes $p_0$-independent, so that
the fermion mass function $M(p_0,P)$ obtained in the self-consistent way
should be {\it $p_0$-independent}, $M(P)$.
Thus the SD equation in the IE approximation reads
\begin{eqnarray}
M(P) =   {\Ga \over 2\Gp}  \int_0^\GL   d Q
{Q M(Q) \over P E_Q}
 \Big[  \tanh {\Gb(E_Q+\Gm) \over 2}
+  \tanh {\Gb(E_Q-\Gm) \over 2} \Big]
 \ln {(P+Q)^2+m^2 \over (P-Q)^2+m^2}.
\end{eqnarray}
In the IE approximation our choice of the photon propagator gives the same
SD equation in the two formalism; ITF and RTF (or TFD).
\par
In the zero-temperature limit (at finite-density $\Gm\not=0$),
the SD equation reads
\begin{eqnarray}
M(P) =   {\Ga \over \Gp}  \int_0^\GL   d Q
{Q \over P E_Q}M(Q) \Gq(E_Q-\Gm)
  \ln {(P+Q)^2+m^2_{T=0} \over (P-Q)^2+m^2_{T=0}}.
\end{eqnarray}

\subsection{numerical results ($\Gm = 0$)}

We have solved the integral equation:
\begin{eqnarray}
M(P) =   {\Ga \over \Gp}  \int_0^\GL   d Q
{Q \over P E_Q}M(Q)
\tanh [{\Gb \over 2} E_Q ]
\ln {(P+Q)^2+m_T^2 \over (P-Q)^2+m_T^2},
\  E_Q = \sqrt{Q^2+M(Q)^2},
\end{eqnarray}
where the photon mass is borrowed from the 1-loop calculation
\cite{kn:Weldon82}:
\begin{eqnarray}
  m_T^2 = {1 \over 3} e^2 T^2 = {4\Gp \over 3} \Ga T^2 .
\end{eqnarray}

In the numerical calculation the dimensionful quantities e.g.,
$T, P, Q, M(P), E_Q, \cdots$ is
normalized by the cutoff $\GL$, and hence
the SD equation can be rewritten in terms of dimensionless variables
where the range of integration is the finite interval $[0,1]$.

First of all, we study the zero-temperature limit $T=0$:
\begin{eqnarray}
M(P) =   {\Ga \over \Gp}  \int_0^\GL   d Q
{Q \over P E_Q}M(Q)
\ln {(P+Q)^2 \over (P-Q)^2},
\  E_Q = \sqrt{Q^2+M(Q)^2} .
\end{eqnarray}
The fermion mass is dynamically generated due to the
nontrivial solution $M(p)\not=0$  and
the chiral symmetry is spontaneously broken in the strong coupling
region $\Ga>\Ga_c$ with the critical coupling $\Ga_c=0.39$.
\par
Fig.3 shows the coupling-constant $\Ga$ dependence of the fermion mass
$M(0)$ which is identified with a pole position of the fermion
propagator: $p^2=M^2(p^2) \simeq M^2(0)$.
Fig.4 exhibits the fermion mass function $M(P)$ which is monotonically
decreasing in $P$.
These features are qualitatively the same as the well-known  results
obtained from the zero-temperature SD equation in QED \cite{kn:Kondo92p}.
\par
At finite temperature $T > 0$, Fig.5 corresponds to Fig.3
at zero-temperature.
The critical coupling $\Ga_c$ depends on the temperature and
$\Ga_c(T)$ is greater than $\Ga_c$ at zero-temperature.
Temperature-dependence of $M(0)$ for a fixed
$\Ga(>\Ga_c)$ is
exhibited in Fig.6(a).  This implies that there exists a critical
temperature $T_c$  above which the chiral symmetry restores in the sense
that the dynamical fermion mass vanishes: $M(0)=0$ for $T>T_c$.

In Fig.7 we compare the fermion mass functions $M(P)$ for various
temperature $T$ at a fixed $\Ga$.
For relatively low-temperature, there exists a peak for $M(P)$.
However the peak disappears as the temperature increases and the critical
temperature is approached.

The phase diagram for two-coupling space $(\Ga, T)$ is depicted in Fig.8
where the low-temperature strong coupling phase where the chiral symmetry
is spontaneously broken
is separated by the critical line from the high-temperature
chiral-symmetry-restoring phase.
This shows that no matter how large the coupling constant $\Ga$ is taken
there exists a finite critical temperature $T_c < \infty$, namely, the
chiral symmetry always restores at a finite temperature $T_c$.

To see the scaling of the dynamical fermion mass,
the dimensionless quantity $M(0)/T_c$ in the region $T/T_c \simeq 1$
is plotted in Fig.9 for various values of
$\Gq:=(\Ga-\Ga_c)/\Ga_c$, $\Ga=\Ga(T\rightarrow 0)=0.36$.
Fig.9 suggests the existence of the
scaling function, i.e., $M(0)/T_c$ is written as a single function of
$T/T_c$ near the critical temperature $T_c$. Our data are  still
insufficient to specify the critical exponent associated with the
finite-temperature transition. More detailed data will be given in the
subsequent paper.

Finally some remarks on the numerical calculation are in order.
In our numerical calculation, we adopt the Double Exponential (DE)
formula to choose sample points. DE formula is efficient when there is
a weak singularity at the end of the interval of integration.
Actually the integral kernel of the SD equation at $T=0$ has a singular
point at $P=Q$. At finite-temperature $T\not=0$,
on the other hand, the  singularity at $P=Q$ disappear due to the
existence of the photon mass $m_T$. Therefore we can calculate the
integrand also at $P=Q$ without difficulty for
high-temperature.   In our numerical calculations we need at least 200
sample points according to DE formula to guarantees the precision of
$10^{-5}$ for the solution $M(P)$.
However we need more sample points to obtain the stable result in the
relatively low-temperature region.
As the number of sample points becomes larger, the low-temperature result
becomes more stable, i.e.,
the rapid change of $M(0)$ at $T \simeq 0$ disappear
even in  the low-temperature region $(T/\GL < 0.01)$
as shown in Fig.6(b).
In order to check the influence of the singularity to a minumum,
we divided the interval [0,1] into subintervals so that each of the
subinterval contains 50 sample points which are chosen according to DE
formula.  For more details, see \cite{kn:KMN92}

\newpage
\section{$p_0$-independent approximation}
\par
In this section we consider another approximation to the SD equation.
In the IE approximation the second self-energy part $\GS_{ph}$ including
the factor $\coth[{\Gb \over 2}q_0]$ has automatically dropped  from the
SD equation and the mass function $M(p_0,P)$ gets $p_0$-independent.
To go beyond the IE approximation, we
must include the effect of the second self-energy part $\GS_{ph}$.
However the integral equation with two arguments:
\begin{equation}
 M(p_0,P) =  \GS_f(p_0,P;T,\Gm) + \GS_{ph}(p_0,P;T),
\end{equation}
is rather difficult
to solve even in the numerical way.
Therefore we consider $p_0$-independent solution $M(P)$
of the SD equation by requiring
\begin{equation}
 {\pd M(q_0,Q) \over \pd q_0} \simeq 0.
\end{equation}
We try to determine the  $p_0$-independent solution
$M(P)$ self-consistently so as to satisfy this condition.
The previous IE approximation is nothing but a sufficient condition for
the solution to be $p_0$-independent.

\subsection{finite-temperature case  ($\Gm=0$)}
 From the $q_0$-independence of the solution, the first self-energy part
reads
\begin{eqnarray}
\GS_f(p_0,P;T,0)
&=&  {\Ga \over 2\Gp} \int_0^\GL   d Q
{Q  \over P E_Q} M(Q)  \tanh {\Gb E_Q \over 2}  \times
\nonumber\\
&&  \times \left[
 \ln {(p_0-E_Q)^2-(E_m^{+})^2 \over (p_0-E_Q)^2-(E_m^{-})^2}
 +
 \ln {(p_0+E_Q)^2-(E_m^{+})^2 \over (p_0+E_Q)^2-(E_m^{-})^2}
\right].
\end{eqnarray}
with
\begin{eqnarray}
E_Q := \sqrt{Q^2+M(Q)^2}.
\end{eqnarray}
Here the value of $p_0$ is specified below.
Note that the right-hand-side of the above equation can be rewritten in
the form which is invariant under the reflection:
$p_0 \rightarrow -p_0$:
\begin{eqnarray}
\GS_f(p_0,P;T,0)
&=&  {\Ga \over 2\Gp} \int_0^\GL   d Q
{Q  \over P E_Q} M(Q)  \tanh {\Gb E_Q \over 2}  \times
\nonumber\\
&& \times
 \ln {[p_0^2-(E_m^{+}+E_Q)^2][p_0^2-(E_m^{+}-E_Q)^2]
 \over
[p_0^2-(E_m^{-}+E_Q)^2][p_0^2-(E_m^{-}-E_Q)^2]}.
\end{eqnarray}
On the other hand, the second self-energy part reads
\begin{eqnarray}
\GS_{ph}(p_0,P;T)
=    {\Ga \over 2\Gp}
\int_0^\GL  dQ {Q \over P E_Q}  M(Q)
  I_\Gb(P,Q,M],
\end{eqnarray}
where
\begin{eqnarray}
I_\Gb(P,Q,M]
&=&   \int_{p_0-E_m^{+}}^{p_0-E_m^{-}} d
q_0
\left( {1 \over q_0+E_Q}-{1 \over q_0-E_Q} \right)
\coth {\Gb(p-q)_0 \over 2}
\nonumber\\
&&+  \int_{p_0+E_m^{-}}^{p_0+E_m^{+}} d q_0
\left( {1 \over q_0+E_Q}-{1 \over q_0-E_Q} \right)
\coth {\Gb(q-p)_0 \over 2} ,
\end{eqnarray}
which is rewritten as
\begin{eqnarray}
I_\Gb(P,Q,M]
&=&
 \int_{E_m^{+}}^{E_m^{-}}
d t \left( {1 \over t-p_0-E_Q}-{1 \over t-p_0+E_Q} \right)
\coth {\Gb t \over 2}
\nonumber\\
&& +  \int_{E_m^{-}}^{E_m^{+}}
d t  \left( {1 \over t+p_0+E_Q}-{1 \over t+p_0-E_Q} \right)
\coth {\Gb t \over 2}
\nonumber\\
&=&   2 \int_{E_m^{-}}^{E_m^{+}}
d q_0
\left[ {q_0+E_Q \over (q_0+E_Q)^2-p_0^2}-{q_0-E_Q \over (q_0-E_Q)^2-p_0^2}
\right] \coth {\Gb q_0 \over 2} .
\end{eqnarray}
Hence $\GS_{ph}(p_0,P;T)$ is also invariant under the replacement:
$p_0 \rightarrow -p_0$.
\footnote{For the $p_0$-independent $M(P)$,
the integration with respect to $q_0$ can be done in principle, but the
result can not be expressed by the elementary function.}
In the zero-temperature limit $(T=0)$,
\begin{eqnarray}
\GS_{ph}(p_0,P;0) &=& {\Ga \over 2\Gp}
\int_0^\GL  dQ {Q \over P E_Q} M(Q)
\nonumber\\
 &&  \ln {[p_0^2-(E_m^{+}+E_Q)^2][p_0^2-(E_m^{-}-E_Q)^2]
\over [p_0^2-(E_m^{-}+E_Q)^2][p_0^2-(E_m^{+}-E_Q)^2]}.
\end{eqnarray}
\par
In the zero-temperature and zero-density limit $(T=0=\Gm)$, therefore,
the SD equation
$M(P)=\GS_{f}(p_0,P;0,0)+\GS_{ph}(p_0,P;0)$ reads
\begin{eqnarray}
M(P) =   {\Ga \over \Gp}
\int_0^\GL  dQ
 {Q \over P}    {M(Q) \over E_Q}
 \ln  {p_0^2-(P+Q+E_Q)^2 \over
 p_0^2-(|P-Q|+E_Q)^2}.
\end{eqnarray}

Now we discuss how to choose $p_0$.
To determine $p_0$, we require
positivity of the nontrivial solution $M(P)$ obtained self-consistently,
at least in the neighborhood $P \simeq 0$, which is satisfied if the
integrand is positive for $P \simeq 0$.  In the zero-temperature limit
$T=0$, this follows from \begin{eqnarray}
p_0 < E_Q+Q .
\end{eqnarray}
In the finite-temperature case $T>0$,
we notice
\begin{eqnarray}
\GS_f(p_0,P;T,0)
&=&  {\Ga \over 2\Gp} \int_0^\GL   d Q
{Q  \over P E_Q} M(Q)  \tanh {\Gb E_Q \over 2}
\nonumber\\
&& \times
 \ln
 \left[1+{4(E_Q+\sqrt{Q^2+m^2})Q P \over
        [(E_Q+\sqrt{Q^2+m^2})^2-p_0^2]\sqrt{Q^2+m^2}}
         + {\cal O}(P^2) \right]
\nonumber\\
&& \times \left[1-{4(E_Q-\sqrt{Q^2+m^2})Q P \over
        [(E_Q-\sqrt{Q^2+m^2})^2-p_0^2]\sqrt{Q^2+m^2}}
        + {\cal O}(P^2) \right],
\end{eqnarray}
where we have used
\begin{eqnarray}
 E_m^{\pm} = \sqrt{Q^2+m^2}  \pm {QP \over \sqrt{Q^2+m^2}}
 + {\cal O}(P^2) .
\end{eqnarray}
In the total (self-consistent) mass function $M(P)=\GS_f+\GS_{ph}$,
we assume the first self-energy part $\GS_f(p_0,P;T,0)$
is the dominant part.
Hence  $\GS_f(p_0,P;T,0)$ has the same signature as $M(P)$,
since it is not difficult to show $p_0$ cannot be chosen
such that both $\GS_f(p_0,P\simeq 0;T,0)$ and
$\GS_{ph}(p_0,P\simeq 0;T)$ are
simultaneously positive (or negative).
Thus
positivity requirement of
both the total mass function $M(P)$ and
the first self-energy part $\GS_f(p_0,P\simeq 0;T,0)$
 restrict the range of $p_0$ as
\begin{eqnarray}
 0 < E_Q-\sqrt{Q^2+m^2} < p_0 < E_Q+\sqrt{Q^2+m^2} .
\end{eqnarray}
Here $E_Q>\sqrt{Q^2+m^2}$, i.e.,
$M(Q)>m=\sqrt{{4\Gp \over 3} \Ga N_f} T$
is satisfied for sufficiently strong coupling $\Ga > \Ga_c \succeq 1$
and sufficiently low-temperature $T/\GL \ll 1$
in the case of finite fermion flavor $N_f$,
while it is always satisfied in the quenched limit $N_f \rightarrow 0$.
This condition does not guarantee
positivity of the second part $\GS_{ph}(p_0,P;T)$ even at $T=0$.
\footnote{The fact that $\GS_f(p_0,P;T)$ and $\GS_{ph}(p_0,P;T)$ has an
opposite sign is consistent with the result of the perturbation theory at
finite temperature \cite{kn:KS85a}.}
The apparently simplest choice $p_0=0$ is excluded by this condition.

\subsection{numerical results ($\Gm = 0$)}
In the numerical calculation we take the simplest choice:
\begin{eqnarray}
p_0 = E_Q ,
\end{eqnarray}
and study the effect of the second part of the self-energy contribution.
For this choice, the SD equation for the mass function reads
$M(P)=\GS_f(E_Q,P;T,0)+\GS_{ph}(E_Q,P;T)$
which can be rewritten in the form:
\begin{eqnarray}
M(P)
=    {2\Ga \over \Gp}
\int_0^\GL  dQ {Q \over PE_Q}  M(Q)
   \int_{E_m^{-}}^{E_m^{+}}
d q_0
{[q_0^2-2E_Q^2] \tanh [{\Gb \over 2}E_Q]-
E_Q q_0 \coth [{\Gb \over 2} q_0]
\over q_0[q_0^2-(2E_Q)^2]},
\end{eqnarray}
where
\begin{eqnarray}
 E_m^{\pm} := \sqrt{| P\pm Q|^2+m_T^2},
 \ m_T^2 = {4\Gp \over 3} \Ga T^2.
\end{eqnarray}
At first, we study the zero-temperature limit.
\begin{eqnarray}
M(P)
=    {\Ga \over \Gp}
\int_0^\GL  dQ {Q \over PE_Q}  M(Q)
 \ln {(P+Q)(P+Q+2E_Q) \over |P-Q|(|P-Q|+2E_Q)}.
\end{eqnarray}
The chiral symmetry is spontaneously broken in the strong coupling region
$\Ga>\Ga_c=0.62$ due to dynamical generation of the fermion mass $M(0)$,
as shown in Fig.10. The dynamical fermion mass function $M(P)$ is shown in
Fig.11. These results are qualitatively in  good agreement with those in
the IE approximation.

Next, we study the finite-temperature case.
Fig.12 shows the temperature-dependence of the fermion mass $M(0)$.
The critical temperature $T_c$ exists  at $T_c \sim 0.0651$,
above which the chiral symmetry restores.

Temperature-dependence of the fermion mass function is shown in Fig.13.
This shows the same tendency as the case under IE approximation.

\subsection{comparison with results in two approximations}

Finally we compare two results obtained in IE approximation
and in $p_0$-independent approximation.

Fig.14 is the plot of the dynamical mass $M(0)$ obtained at T=0 under
two approximations where the horizontal axis is the coupling constant
normalized respectively by each critical coupling.
This shows that $M(0)$ under $p_0$-indep. approximation is slightly
smaller than IE approximation for all over the range of the coupling
constant.  In other words, additional effects from $\GS_{ph}$ work to
decrease $M(P)$ at T=0.

Fig.15 shows the  plot of the $M(0)/T_c$ versus $T/T_c$ for the same
ratio $\Ga_c(T=0)/\Ga=0.618$.
Hence $M(0)/T_c$ in IE approximation exhibits larger value for $T/T_c
\simeq 0$ and smaller value for $T/T_c \simeq 1$ than $p_0$-indep.
approximation.

Concerning the transition temperature, IE approximation gives
slightly smaller critical temperature $(T_c/\GL=0.063)$  than $p_0$-indep.
approximation $(T_c/\GL=0.065)$. This implies that
effect of the retarded propagation of the photon has a tendency of
raising the critical temperature.

\newpage
\section{Conclusion and Discussion}
As the formalism which can treat the field theory at
finite-temperature and finite-density, we know two formalism:
Imaginary-Time Formalism (ITF) and Real-Time Formalism (RTF).
In this paper we have derived the Schwinger-Dyson (SD) equation
at finite-temperature and non-zero density as a
tool to analyze the non-perturbative effect
in the Nambu-Jona-Lasinio (NJL) model of four-fermion interaction  and
QED.
In section 3, we have shown that two formalism give the same SD
equation for the  NJL model in the leading order of 1/N expansion, i.e.,
chain approximation.
\par
In D-dimensional QED (QED$_D$) at finite-temperature and nonzero-density
we have derived the SD equation for the fermion propagator under the bare
vertex ansatz (approximation).
This is a coupled integral equation
for the fermion mass function $M(p_0, P)$ and the wave function
renormalization function $Z(p_0, P)$ with two-variables
$p_0, P=|\vec P|$.
\par
In order to obtain the explicit solution,
we have put $Z(p)\equiv 1$ from the outset and searched for
the {\sl real} solution of the SD equation for the fermion mass function
$M(p)$ in QED$_4$ at finite-temperature and nonzero-density.
Even in this stage, solving the two-variable integral equation
is rather difficult even in the numerical calculation.
Therefore we consider the approximation so that the SD equation reduces
to the one-variable integral equation whose solution is denoted by $M(P)$.
Actually we have adopted two types of approximations:
instantaneous-exchange (IE) approximation and the $p_0$-independent
approximation. The IE approximation in TFD adapted in this paper leads to
the same SD equation as that derived  under the corresponding IE
approximation in ITF \cite{kn:DM91a}. In IE approximation only one
temperature-dependent factor $\tanh[{\Gb \over 2}E_P]$ appears in the SD
gap equation and hence all the effects of temperature comes from this
factor. Moreover IE approximation greatly simplifies the gap equation,
since IE approximation automatically render the solution $M(p)$
$p_0$-independent: $M(P)$. \par
The SD equation derived in this paper based on RTF
immediately allows us to do another approximation, i.e.,
$p_0$-indep. approximation which goes beyond IE approximation. This
approximation requires, from the first, that the
solution $M(p)$ is  $p_0$-independent. However this approximation is able
to incorporate a new temperature-effect coming from the factor $\coth
[{\Gb \over 2}q_0]$, which is disregarded in IE approximation.
\par
Numerical calculations under the two approximations have shown
qualitatively the same result.
For any coupling constant $\Ga(>\Ga_c)$ belonging to the strong coupling
region where the chiral symmetry is spontaneously broken in the
zero-temperature, there exists a finite critical temperature
$T_c(\Ga) < \infty$
above which the chiral symmetry restores.
No matter how the coupling $\Ga$ may be strong, the critical temperature
$T_c$ is kept finite.

\par
In this paper we have not analyzed the finite-density transition.  In the
presence of chemical potential $\Gm\not=0$, the symmetry $p_0 \rightarrow
-p_0$ will be lost in the SD equation and we are forced to tackle with the
original two-variable integral equation.
Therefore approximations adopted in this
paper must be improved to treat such a case.  This will be a subject in
the forthcoming paper.

\section{Acknowledgments}
The authors would like to thank Tadahiko Kimura and Toru Ebihara
for critical comments and Takashi Heya for help in gathering the data of
numerical calculations.
Numerical calculations were done on HITAC S-820/80
and HITAC S-3800/480 at
Computation Center of University of Tokyo.

\newpage
\section{Figure Captions}
\begin{enumerate}
\item[Fig.1:]
Fermion self-energy contribution in the NJL model.

\item[Fig.2:]
Fermion self-energy contribution in QED.

\item[Fig.3:]
$M(0)$ vs $1/\Ga$ at $T = 0$ in IE approximation, ($\Ga_c = 0.39$).

\item[Fig.4:]
Fermion mass function at $T = 0$ in IE approximation, ($\Ga = 0.40$).

\item[Fig.5:]
$M(0)/\GL$ vs $1/\Ga$ at $T/\GL = 0.0794$ in IE approximation,
($\Ga_c = 0.70$).

\item[Fig.6:]
(a) Temperature-dependence of the fermion mass in IE approximation
 ($\Ga = 0.63$, $T_c/\GL = 0.0628$),
(b) Dependence of $M(0)$ on the number of sample points $N$
in the low-temperature region $T/\GL < 0.01$.

\item[Fig.7:]
Temperature-dependence of the fermion mass function in IE approximation
for
$T/\GL=10^{-9}, 10^{-7}, 10^{-5}, 10^{-3}, 0.04$ ($\Ga=0.70$).

\item[Fig.8:]
Phase diagram of QED$_4$ at finite-temperature in IE approximation.

\item[Fig.9:]
$M(0)/T_c$ vs $T/T_c$ in IE approximation where
$\Gq = 0.08, 0.75, 0.94, 1.7.$

\item[Fig.10:]
$M(0)/\GL$ vs $1/\Ga$ at $T = 0$ in $p_0$-indep. approximation,
($\Ga_c = 0.618$).

\item[Fig.11:]
Fermion mass function at $T = 0$ in $p_0$-indep. approximation,
($\Ga = 1.0$).

\item[Fig.12:]
Temperature-dependence of the fermion mass in $p_0$-indep. approximation
 ($\Ga = 1.0$, $T_c/\GL = 0.0651$).

\item[Fig.13:]
Temperature-dependence of the fermion mass function
in $p_0$-indep. approximation for
$T/\GL=10^{-5}, 10^{-3}, 0.024$ ($\Ga=1.0$).

\item[Fig.14:]
Comparison of IE and $p_0$-indep. approximations:
$M(0)/\GL$ vs $\Ga_c/\Ga$.

\item[Fig.15:]
Comparison of IE and $p_0$-indep. approximations:
$M(0)/T_c$ vs $T/T_c$ for $\Ga_c/\Ga=0.618$.

\end{enumerate}

\end{document}